# 120-fs single-pulse generation from stretched-pulse fiber Kerr resonators

Xue Dong,* Zhiqiang Wang, and William H. Renninger

*Institute of Optics, University of Rochester, Rochester, New York 14627, USA*
*Corresponding author: xdong18@ur.rochester.edu*

**Fiber Kerr resonators are simple driven resonators with desirable wavelength and repetition rate flexibility for generating ultrashort pulses for applications including telecommunications, biomedicine, and materials processing. However, fiber Kerr resonators to date often generate longer pulses and require more complicated techniques for generating single pulses than would be desirable for applications. Here we address these limits by demonstrating robust single-pulse performance supporting 120-fs pulse durations in fiber Kerr resonators based on stretched-pulse solitons. Through matching numerical and experimental studies, stretched-pulse soliton performance is found to strongly depend on the total cavity length, and the optimum length is found to depend on the drive, Raman scattering, and the total pulse stretching. The bandwidth increases with decreasing net dispersion, enabled by shorter total cavity lengths. In a cavity with an optimized length and the described setup, stable stretched-pulse solitons corresponding to 120-fs duration are experimentally observed. In addition, soliton trapping is demonstrated with a pulsed drive source despite large intracavity breathing and single-pulse performance is observed. Robust with high performance single-pulse generation is a critical step toward useful femtosecond pulse generation.**

Ultrashort pulse mode-locked lasers are essential for applications including telecommunications, biomedicine, and materials processing. Fiber-based mode-locked lasers are robust and cheap, enable low repetition rates, and rival the performance of bulk systems including those featuring Ti:Sapphire gain media. The essential mechanism for ultrashort pulse generation in a mode-locked laser is soliton formation, with major performance advances coming with novel solitons including those associated with stretched pulse, chirped pulse, self-similar, and Mamyshev mode-locking [1–4]. However, despite these incredible performance advances, the requirement of a laser gain medium places fundamental limits on the laser wavelength, requiring complex and expensive systems for generating wavelengths not convenient to the laser emission band. Efficient operation of the gain medium also requires a minimum interaction region, which limits the repetition rate and ultimately the energy of the pulses.

Kerr resonators are an emerging alternative for ultrashort pulse generation [5]. Kerr resonators are a flexible source of coherent pulses because they circumvent the issues associated with a gain medium, because instead of gain, they compensate for loss with an external coherent continuous-wave drive. Kerr resonator frequency comb generation is also based on soliton formation, with behavior related to that in mode-locked lasers. Kerr resonators have developed considerably in the micro-scale on chip, enabling GHz to THz pulse repetition rates and low threshold powers, and establishing applications in spectroscopy, distance measurements, and frequency synthesis [6,7]. In fiber, researchers have focused on all-optical buffering, long range interaction, and temporal tweezing [5,8,9]. However, before Kerr resonators can find widespread use for traditional mode-locked laser applications, they face many challenges, including low peak powers [5] and difficulty in maintaining a single-soliton state [10,11].

Pulse energy and duration are crucial performance parameters for many applications. The pulse performance of fiber Kerr resonators promises improvement with the recent discoveries of advanced solitons such as stretched pulse solitons [12,13], chirped pulse solitons [14,15], higher-order dispersion enabled solitons [16], and Raman-assisted solitons [17]. For example, while early measured fiber Kerr solitons were ps in duration, including 4 ps in 2010 [5] and 2.6 ps in 2013 [8], a more recent study found 230-fs pulses using a novel generation technique based on third order dispersion [16]. However, fiber Kerr resonator solitons are still longer than the durations obtained from high-performance mode-locked fiber lasers. In a notable recent experiment featuring a careful balance of small normal dispersion, temporal desynchronization, strong Raman gain, and third-order dispersion, a broad spectrum was observed from a Kerr resonator, but without temporal measurement [17]. A general and simple technique for stable short pulse generation in fiber Kerr resonators is needed. In dispersion-managed mode-locked laser systems the stretched-pulse soliton is well known to support stable short pulses [3]. For Kerr resonators, early research on dispersion-managed cavities focused on parametric instabilities, mechanisms for temporal binding, and the emission of dispersive waves [18–20]. Recently, 210-fs duration stretched-pulse solitons were observed in fiber cavities with a net dispersion close to zero [12]. However, this pulse duration remains longer than what is observed from stretched pulse mode-locked lasers, which suggests that shorter pulses are achievable from stretched-pulsed soliton Kerr resonators.

For ultrashort pulse applications, a single reproducible soliton is generally preferred. A freely running Kerr resonator driven by a continuous-wave laser generates solitons stochastically and it can be challenging to obtain a single-soliton state. Several methods have been demonstrated to address the challenge and obtain single solitons including phase-modulated pumping and backward laser tuning [10,11]. Soliton trapping is an effective technique for single pulse generation which has now been demonstrated in microresonators [21], bulk enhancement cavities [22], as well as fiber cavities [23]. However, these studies focused on traditional solitons with negligible intracavity dynamics. It is not clear if similar techniques can be applied to solitons that have large intracavity dynamics such as stretched-pulse solitons.

Here we investigate the limits of pulse performance and control for stretched-pulse fiber Kerr resonators. By investigating the interplay of nonlinearity, Raman scattering, dispersion-management, drive power, and pulse stretching, we find that for fixed cavity lengths the bandwidth is limited by the drive power, the Raman effect and the soliton stretching ratio. These limitations, moreover, have a strong dependence on the total cavity length. Here we optimize the cavity within these constraints to generate and characterize stable stretched-pulse solitons with 120-fs duration and present guidelines for further improvement. Experimental observations are in excellent agreement with numerical predictions. In addition, using this accurate model for guidance, we demonstrate single-pulse soliton trapping of ultrashort stretched-pulse solitons for the first time. This enables controllable singe-pulsing at the highest performances levels. The general working principles established, and the performance and control demonstrated will be valuable for practical wavelength-versatile femtosecond pulse generation from Kerr resonators.

Stretched-pulse solitons are generated in cavities featuring alternating sections of anomalous and normal dispersion fibers. These Gaussian-shaped solitons are referred to as stretched pulses because they periodically stretch and compress throughout the dispersion-managed cavity. Stretched pulsed solitons were initially developed in mode-locked fiber lasers, where they were shown to support shorter pulse durations than solitons based on traditional all-anomalous dispersion systems [3]. Stretched pulsed solitons have also been studied theoretically and experimentally in Kerr resonators [12,13]. In Kerr resonators, stretched-pulsed solitons feature similar Gaussian profiles and stretching dynamics. The spectral bandwidth and corresponding ultrashort pulse duration were found to have a strong dependence on the total cavity dispersion and drive power, with the shortest pulses found at the largest available powers with a total dispersion near zero [12]. In Ref. [12], it was also found that the spectral bandwidth has some dependence on the total length of the cavity. However, because the lowest measurable length was limited by the available drive power, the performance limitations and exact dependence on cavity length remain unclear. Therefore, in this study, with access to larger drive powers, we examine the stretched-pulse soliton's dependence on total cavity length and investigate higher performance designs.

Dispersion-managed fiber cavities are designed to study the dependence of stretched-pulse solitons on the total cavity length (Fig. 1a). Each cavity consists of two commercially available fibers with opposite signs of group velocity dispersion and varying lengths [12] (smf28 has group-velocity dispersion (GVD) -23 $ps^2$/km, third-order dispersion (TOD) 0.1 $ps^3$/km and nonlinear coefficient 1.3 /W/km and metrocor has GVD 9.7 $ps^2$/km, TOD 0.1 $ps^3$/km and nonlinear coefficient 2.1 /W/km). A drive laser is frequency locked with respect to a resonant frequency of the cavity with a proportional-integral-derivative control circuit. The cavity phase detuning is controlled through the frequency offset from the cavity resonance. The drive laser intensity is modulated with a fiber-format electro-optic modulator into a train of 0.3-ns pulses with a repetition rate matched to the cavity's. The modulated drive is amplified up to an average power of 0.6 W with an erbium-doped fiber amplifier before being input to the cavity through a 5% fiber coupler. The cavity output is measured from an additional 2% fiber coupler. To study the dependence on the cavity parameters, the cavity is addressed using 2-ps pulses from an amplified mode-locked laser. The addressing pulses are gated with an intensity modulator to selectively excite stretched-pulse solitons through cross-phase modulation (Fig. 1a and Supplement, Section 6). At each total cavity length, the net dispersion is varied from large and negative to zero by appropriately adjusting the length of the two fibers. The bandwidth is known to increase with drive power [12], with an upper limit determined by either drive availability or the onset of Raman instabilities [24,25]. Experimentally, Raman instabilities are inferred from initial generation of new frequencies a characteristic Raman shift from the drive wavelength-at higher wavelengths (additional details in Supplement 1, Section 1). The experimental max drive power increases with decreasing cavity length as 42/L[m] W (see Supplement 1, Section 3). This inverse dependence with length stems from the linear dependence of the Raman effect on length. For each cavity the detuning is optimized for the broadest bandwidth and the spectra are recorded at the highest power before the onset of Raman instabilities. This is repeated as the net dispersion is reduced toward zero until stable solitons are no longer observed. In summary, for each total length, the net dispersion, detuning and drive are optimized for the broadest spectrum. The spectral bandwidth at this optimum is found to increase with decreasing total cavity length (Fig 1(b)-(c)).

The stretched-pulse soliton's dependence on the total cavity length is also investigated numerically. A stretched-pulse fiber Kerr resonator can be accurately modeled using an Ikeda-type model [12] where the fiber dispersion and nonlinearity are given by the manufacturer (listed above) and the total loss is fixed by the roundtrip component loss (16%). At a given total cavity length, stable stretched-pulse solitons are obtained at the Raman-limited drive power determined experimentally for different detuning and group delay dispersion (GDD) (relative lengths of the two fiber sections) values, and the solution with the broadest bandwidth is selected. The precision of the detuning is important, as detailed in Supplement 1, Section 3. In excellent agreement with experimental observations, broader bandwidth solitons are stable with cavities with shorter total lengths (blue line in Fig 1(c)). These simulations are repeated without the TOD of the fiber, using the same drive limit and with reoptimized detuning and net GDD values. The full-width half maximum spectral bandwidth is found to increase at every cavity length without TOD (red line in Fig. 1(c)). Notably, however, because of the growth of a low-wavelength TOD-enabled dispersive wave, the root-mean-squared bandwidth can still be larger with TOD (see Supplement 1, Section 3).

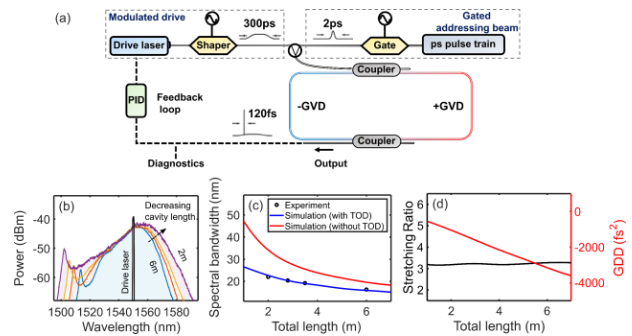

Fig. 1. (a) Schematic of the experimental setup. (b) Broadest measured spectrum as a function of total cavity length. The drive laser frequency is plotted in black. (c) The broadest spectral bandwidth at full-width-half-max vs. total cavity length from experiments (points) and simulations with (blue line) and without (red line) TOD. (d) The smallest net dispersion that supports stretched pulses vs. total cavity length (red) and the pulse stretching ratio at the smallest net dispersion vs. total cavity length (black).

To gain insight into the stretched-pulse soliton dependence on total length, we numerically find the GDD supporting the broadest bandwidth (from Fig. 1(d)) vs. length. Without TOD (for simplicity; see Supplement 1, Section 3 for analysis with TOD), shorter cavities yield the broadest bandwidths at smaller magnitudes of net dispersion (Fig 1(d)). Broader bandwidths at shorter lengths are consistent with the known inverse dependence of bandwidth with net dispersion for stretched pulse solitons. However, it is unclear what limits the net dispersion for each length. To further investigate this limit, we examine the stretching ratio (max/min duration) of the solitons for the optimized bandwidth solutions from Fig. 1(d). Interestingly, the stretching ratio is found to be constant with total length (Fig 1(d)). This implies that the decrease in stretching from the decreasing cavity length is balanced by the increase in stretching from the increase in bandwidth from the decreasing net dispersion. Therefore, given an upper limit of the stretching ratio that can be stabilized, pulses can be stabilized at a smaller net dispersion and a broader bandwidth when the cavity total length is small. In summary, the bandwidth increases because the net dispersion is decreased, which is enabled by decreasing the total cavity length.

From the theoretical analysis above, the shortest pulses can be obtained by reducing the total cavity length. However, with shorter cavities the total nonlinearity decreases, which increases the drive power that will be required by a comparable amount. This drive dependence on nonlinearity is found from a normalized field equation analysis [15] or the parametric threshold power for fixed cavity loss. In addition, because the drive originates from an intensity-modulated laser that is amplified to a fixed power, the drive pulse energy (and therefore peak drive power) scales linearly with the round-trip time of the cavity, and therefore its length. Overall, with decreasing length the drive requirement increases linearly, and the available drive decreases linearly. The highest bandwidth result is obtained at the shortest cavity length for which the maximum available drive coincides with the Raman instability threshold. This limit is found with a 2-m total cavity length. Above this maximum 18-W drive, Raman instabilities are observed, and for shorter cavities the drive power is too low to achieve broader bandwidths. The broadband 22-nm spectrum observed (Fig. 2(a) black) corresponds to a transform-limited pulse with 120-fs duration. Numerical simulations of stable stretched pulse solitons in a cavity matching the experimental parameters are in excellent agreement with experimental results (Fig. 2(a) blue). Note that the slight deviation at long wavelengths may arise because the simulation does not include the Raman effect which is known to upshift the cavity soliton central wavelength [24].

The output pulses are also characterized temporally with autocorrelation measurements. After the cavity, the low frequency drive is spectrally filtered out with a fiber-Bragg-grating filter and the pulses are amplified by an Er-doped fiber amplifier. The fibers between the cavity and the autocorrelator have a total of ~46 second order and ~5 third order dispersion lengths, which would yield pulses much longer than 120 fs. By adding 1-m additional dispersion compensating fiber, the total fiber contributes no residual second order dispersion and 2 residual third order dispersion lengths. With this residual third order dispersion, the measured autocorrelation corresponds to pulses with 150-fs duration (Fig 2 (b) black). The excellent agreement with simulations (Fig 2 (b) blue) confirms the generation of 120-fs pulses directly from the fiber Kerr resonator (see Supplement 1, Section 4).

In the measurements above, the number of pulses is not controlled. Stable measurements are obtained using the widely adopted technique of continuously addressing the cavity with an auxiliary source [5,8,12]. However, practical sources for ultrashort-pulse applications often require a single stable pulse.

A single stable pulse has been shown to be achievable using a pulse trapping technique, through which a drive pulse traps a single soliton when the desynchronization between the drive repetition rate and cavity repetition rate is small [23]. This method has been experimentally verified in fiber, on chip, as well as bulk enhancement cavities [21–23]. However, these previous results focused exclusively on traditional solitons with negligible intracavity dynamics. It is not clear if the trapping technique can be applied to ultrashort stretched-pulse solitons.

Stretched-pulse soliton trapping is investigated numerically using the system parameters from above. The drive is defined as a fourth-order super-Gaussian pulse with 0.3-ns pulse duration and 18-W peak power, in contrast to the 18-W continuous-wave drive used above (see Supplement 1, Section 5). The desynchronization between the drive and the cavity is implemented through a constant relative velocity applied to the drive pulse. In absence of nonlinear trapping, this relative velocity forces the drive pulse to drift away from the stable soliton. Trapping is therefore evidenced by a lack of numerical drift. We find that for stretched-pulse solitons, if the desynchronization is sufficiently small (e.g. 0.005 fs per round trip), the trapping force compensates for the drift, and trapping occurs (Fig 3(a)). See Supplement 1, Section 5 for additional theoretical trapping details and analysis.

Stretched-pulse soliton trapping is investigated experimentally using the setup described above. The number of solitons excited in the cavity is controlled by gating the input addressing pulse train with an intensity modulator (Fig. 1(a)). The drive pulse repetition rate is set close to the cavity's (as measured by the cavity pulse period) with a mismatch of <0.1 Hz. Under this condition, a single pulse is stabilized at the edge of the pump pulse (Fig 3(b)). This experimental trapping threshold agrees well with theoretical predictions (0.005ft/rt corresponds to 0.05 Hz for the 2-m cavity). After initiation, no further addressing pulses are required to maintain the soliton in the cavity. Single pulse operation is verified with the combination of direct fast detection (40-ps resolution) and optical spectrum measurements with 0.12-nm resolution.

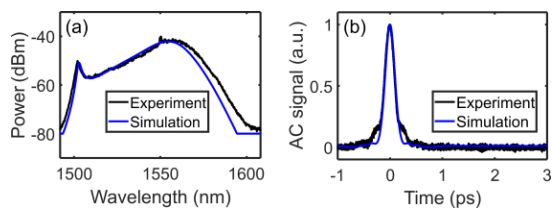

Fig. 2. Shortest stretched-pulse soliton (a) measured (black) and simulated (blue) spectrum, (b) measured (black) and simulated (blue) compressed output autocorrelation.

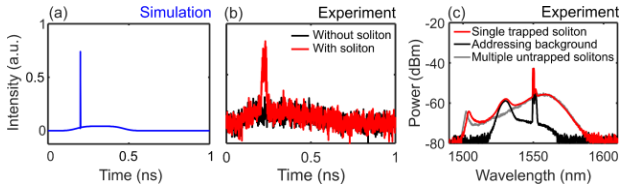

Fig. 3. Stretched-pulse soliton trapping. (a) Numerically simulated pulse for drive-to-soliton desynchronization +0.005fs/rt. (b) Experimental temporal profile of the cavity output with and without a single stretched-pulse soliton. (c) Experimental output spectrum of a single trapped soliton (red), of the addressing source background without a soliton (black), and of multiple un-trapped solitons from Fig. 2a normalized for comparison.

The spectrum with one trapped soliton (Fig. 3c red) is weaker than the un-trapped solitons that are continuously seeded (Fig. 3c grey), because there are fewer pulses. Because the soliton spectrum is low power, an additional peak at 1530nm from the spontaneous emission noise of the seed amplifier can also be observed (see black line in Fig. 3c and Supplement 1, Section 6). Importantly, the trapped single soliton component of the spectrum (Fig. 3c red) agrees with the spectrum for the un-trapped solitons in the continuously seeded case from Fig. 2 (Fig. 3c grey), confirming the stability of single ultrashort stretched pulse solitons.

Here, stretched-pulse fiber Kerr resonators were found to support broader bandwidths and shorter pulses when the total cavity length is reduced. As discussed above, further length reductions are limited by available drive power. Therefore, shorter pulses are immediately accessible, using the design guidelines developed in this study, with higher peak drive powers. While higher average powers will be limited by damage to optical components, the peak drive power can also be increased with fixed average power with shorter drive pulses. Shorter drive pulses have the additional benefit of increasing the drive-to-pulse efficiency because there is a higher overlap of the drive with the solitons in time. Shorter pulse driving, however, adds additional system cost.

With shorter pulses, higher-order effects will become more relevant. For example, Raman scattering is known to place a fundamental limit on the pulse duration for traditional anomalous dispersion Kerr resonator solitons [24]. While the results from [24] do not strictly apply for the small net-dispersion values here, it would be valuable to theoretically investigate the impact of Raman scattering on stretched-pulse solitons as well. Experimentally, it may be possible to change or mitigate the effects of Raman scattering using for example different fibers or by employing spectral filtering. Further investigation into the effects of Raman scattering is particularly valuable for stretched-pulse solitons because as detailed above, it sets the first limit to drive power and therefore bandwidth for longer cavities.

In summary, we present new critical parameter dependences and limitations to stretched-pulse solitons and apply these results to obtain 120-fs pulses. Single pulsing is achieved through the first demonstration of stretched-pulse soliton trapping. These results motivate further development of stretched-pulse soliton Kerr resonators for application-focused ultrashort pulse generation.

**Funding.** National Institute of Biomedical Imaging and Bioengineering (R01EB028933).

**Disclosures.** The authors declare no conflicts of interest.
**Supplemental document**. See Supplement 1 for supporting content.

**REFERENCES**

1. A. Chong, L. G. Wright, and F. W. Wise, Reports Prog. Phys. **78**, 113901 (2015).
2. Z. Liu, Z. M. Ziegler, L. G. Wright, and F. W. Wise, Optica **4**, 649 (2017).
3. S. K. Turitsyn, B. G. Bale, and M. P. Fedoruk, Phys. Rep. **521**, 135 (2012).
4. F. W. Wise, A. Chong, and W. H. Renninger, Laser Photonics Rev. **2**, 58 (2008).
5. F. Leo, S. Coen, P. Kockaert, S. P. Gorza, P. Emplit, and M. Haelterman, Nat. Photonics **4**, 471 (2010).
6. T. J. Kippenberg, A. L. Gaeta, M. Lipson, and M. L. Gorodetsky, Science **361**, eaan8083 (2018).
7. A. Pasquazi, M. Peccianti, L. Razzari, D. J. Moss, S. Coen, M. Erkintalo, Y. K. Chembo, T. Hansson, S. Wabnitz, P. Del'Haye, X. Xue, A. M. Weiner, and R. Morandotti, Phys. Rep. **729**, 1 (2018).
8. J. K. Jang, M. Erkintalo, S. G. Murdoch, and S. Coen, Nat. Photonics **7**, 657 (2013).
9. J. K. Jang, M. Erkintalo, S. Coen, and S. G. Murdoch, Nat. Commun. **6**, 7370 (2015).
10. H. Guo, M. Karpov, E. Lucas, A. Kordts, M. H. P. Pfeiffer, V. Brasch, G. Lihachev, V. E. Lobanov, M. L. Gorodetsky, and T. J. Kippenberg, Nat. Phys. **13**, 94 (2017).
11. D. C. Cole, J. R. Stone, M. Erkintalo, K. Y. Yang, X. Yi, K. J. Vahala, and S. B. Papp, Optica **5**, 1304 (2018).
12. X. Dong, Q. Yang, C. Spiess, V. G. Bucklew, and W. H. Renninger, Phys. Rev. Lett. **125**, 33902 (2020).
13. C. Bao and C. Yang, Phys. Rev. A **92**, 023802 (2015).
14. X. Dong, C. Spiess, V. G. Bucklew, and W. H. Renninger, Phys. Rev. Res. **3**, 033252 (2021).
15. C. Spiess, Q. Yang, X. Dong, V. Bucklew, and W. Renninger, Optica **8**, 861 (2021).
16. Z. Li, Y. Xu, S. Coen, S. Murdoch, and M. Erkintalo, Optica **7**, 1195 (2020).
17. Y. Xu, A. Sharples, J. Fatome, S. Coen, M. Erkintalo, and S. G. Murdoch, Opt. Lett. **46**, 512 (2021).
18. A. U. Nielsen, B. Garbin, S. Coen, S. G. Murdoch, and M. Erkintalo, APL Photonics **3**, 120804 (2018).
19. Y. Wang, F. Leo, J. Fatome, M. Erkintalo, S. G. Murdoch, and S. Coen, Optica **4**, 855 (2017).
20. F. Copie, M. Conforti, A. Kudlinski, A. Mussot, and S. Trillo, Phys. Rev. Lett. **116**, 143901 (2016).
21. E. Obrzud, S. Lecomte, and T. Herr, Nat. Photonics **11**, 600 (2017).
22. N. Lilienfein, C. Hofer, M. Högner, T. Saule, M. Trubetskov, V. Pervak, E. Fill, C. Riek, A. Leitenstorfer, J. Limpert, F. Krausz, and I. Pupeza, Nat. Photonics **13**, 214 (2019).
23. M. Erkintalo, S. G. Murdoch, and S. Coen, J. R. Soc. New Zeal. 1 (2021).
24. Y. Wang, M. Anderson, S. Coen, S. G. Murdoch, and M. Erkintalo, Phys. Rev. Lett. **120**, 053902 (2018).
25. C. Aguergaray, A. Runge, M. Erkintalo, and N. G. R. Broderick, Opt. Lett. **38**, 2644 (2013).

# Supplementary Information -- 120-fs single-pulse generation from stretched-pulse fiber Kerr resonators

Xue Dong, Zhiqiang Wang, and William H. Renninger
*Institute of optics, University of Rochester, Rochester, New York 14627*

## 1. Experimental parameters

Four cavities with different total length are investigated and optimized in the experiment (parameters listed in Table 1). The broadest spectrum for each cavity is shown in Fig 1 in the main paper and the corresponding bandwidth is listed in the fourth column of Table 1. The bandwidth increases by increasing the net dispersion toward zero or increasing the pump power. The optimum bandwidth is achieved for each cavity length at the maximum pump power and smallest magnitude of net dispersion for which pulses are stable. The simulated intracavity pulse energy for each cavity is listed in the last column of Table 1. The pulse energy increases with decreasing cavity length because the pulse duration decreases sublinearly, which is slower than the linear increase of peak power when the cavity length is decreased. The highest simulated pulse energy of 37 pJ from the 2-m cavity agrees well with the experimental energy of 32 pJ estimated from spectral and average power measurements. In summary, as detailed in the paper, net dispersions closer to zero and higher drive powers can be achieved at lower cavity lengths, where the performance is highest.

Table 1. Experimental parameters and results

| Length (m) | Net dispersion (ps^2) | Pump power (W) | Bandwidth (nm) | Pulse energy (pJ) |
|---|---|---|---|---|
| 6 | -0.0052 | 9 | 16.3 | 25 |
| 3.5 | -0.0033 | 14 | 19.2 | 31 |
| 2.8 | -0.0029 | 17 | 20.4 | 35 |
| 2 | -0.0021 | 18 | 22 | 37 |

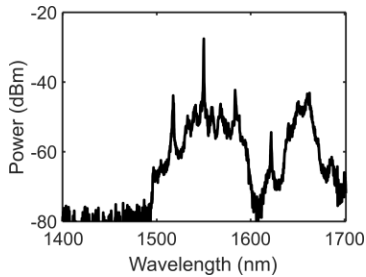

Fig. S1. Measured spectrum for the 2-m cavity for a high drive power of 20W, showing Raman Stokes breakthrough destabilizing soliton generation.

The pump power in the table is the max drive power that can stabilize stretched pulse solitons at each cavity length. Above the listed pump power, a new band is generated at a higher wavelength (e.g. Fig S1 for the 2-m cavity). Raman scattering is known to limit fiber [1] and micro [2] Kerr resonators as well as mode-locked lasers [3]. Specifically, the generation of a distinct Stokes frequency band has been shown previously in mode-locked lasers to destabilize soliton formation [3].

## 2. Numerical modeling

Fiber Kerr resonators can be accurately modeled with an Ikeda-type system of equations including an equation describing the pulse propagation in the waveguide and a periodic boundary condition accounting for the discrete application of the drive and losses. Note that such an Ikeda type model is typically approximated by a Lugiato-Lefever Equation [4–6]. However, the Lugiato-Lefever Equation cannot be used to model the stretched-pulse soliton owing to the large changes per round trip inherent in the pulse dynamics. Pulse propagation in the resonator waveguide, including the relative phase detuning, is modeled by a detuned nonlinear Schrödinger equation. Since the stretched pulse is generated in a dispersion-managed cavity which consists of two types of fiber (SMF-28 and Metrocor), these two fiber segments are modeled separately as:

$$\frac{\partial A}{\partial z} = -\frac{\alpha}{2}A - i\frac{\beta_{2\text{SMF}}}{2}\frac{\partial^2 A}{\partial t^2} + \frac{\beta_3}{6}\frac{\partial^3 A}{\partial t^3} + i\gamma_{SMF}|A|^2 A - i\delta A, \text{ and}$$

$$\frac{\partial A}{\partial z} = -\frac{\alpha}{2}A - i\frac{\beta_{2\text{Metrocor}}}{2}\frac{\partial^2 A}{\partial t^2} + \frac{\beta_3}{6}\frac{\partial^3 A}{\partial t^3} + i\gamma_{Metrocor}|A|^2 A - i\delta A.$$

The periodic boundary condition incorporating the drive, and loss is described as

$$A^{n+1}(0,t) = \sqrt{TD} + A^n e^{-\alpha_0}.$$

These two equations are not known to be reducible into a single equation. Note that as a consequence the stretched pulse has a Gaussian profile and a distinct stretching behavior which is noticeably distinguished from the static hyperbolic secant profile in traditional Kerr resonator systems.

## 3. Additional analysis on the cavity length dependence

In the experiments detailed in the paper, the net dispersion, detuning, and drive are optimized for generating the broadest spectrum. The experimental max drive power increases with decreasing cavity length as 42/L[m] W (Fig S2a). Using this maximum drive power dependence, the broadest spectrum is determined numerically by optimizing the net dispersion and detuning as a function of total cavity length. The addition of third order dispersion (TOD) was found (Fig 1 in the main paper) to reduce the spectral bandwidth as defined by its full-width at half maximum (FWHM). However, TOD also induces a resonant dispersive wave sideband at lower wavelengths which results in an additional increase in the bandwidth as defined by its root-mean squared (RMS). As a result of the increase from the sideband and the decrease of the FWHM, the RMS bandwidth of stretched-pulsed solitons is comparable with and without TOD for each cavity length (Fig S2b), indicating that the transform-limited pulse duration is also comparable with and without TOD.

In the paper, the cavity GDD that supports the broadest bandwidth was determined as a function of total cavity length for the case without TOD. Using the optimum GDD for each cavity length, the stretching ratio was found to be constant as a function of cavity length. When including TOD, the optimum GDD value will be different. In addition, because TOD can induce temporal oscillations which can make the stretching ratio ambiguous, here we examine the number of second and third order dispersion lengths that the pulse propagates through as a function of total cavity length. Note that without TOD, this metric gives a similar result to the stretching ratio (red line in Fig S2c). With TOD, the number of second order dispersion lengths decreases slightly with total cavity length. Interestingly, the number of third order dispersion lengths increases with decreasing cavity length (Fig. S2d), which indicates that the effect of TOD becomes stronger when the cavity length is shorter.

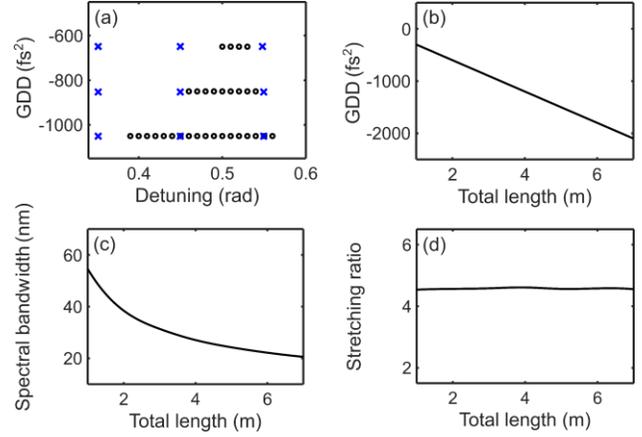

Fig. S3. (a) Location of stable stretched pulse solitons in net dispersion-detuning parameter space with 0.01 rad detuning resolution (black dots). With a coarser 0.1 rad detuning resolution (blue crosses), stable solutions would be missed. (b) Smallest net dispersion that stretched pulses are stable vs. total cavity length. (c) The broadest spectral bandwidth at FWHM vs. total length from simulations. (d) The stretching ratio at the smallest net dispersion as a function of the total cavity length.

The stretched-pulse bandwidth was found to increase with decreasing total cavity length. This bandwidth also depends on the precision with which the detuning is scanned. Note that the detuning is controlled by an external PID controller. The experimental detuning precision is determined by the locking resolution of the PID controller. For a specific cavity and drive power, the stretched pulse is stable over a range of detuning values that shrinks with a decreasing magnitude of GDD (Fig S3a). If the precision of the detuning scan is too course, stable solutions with smaller GDD and hence broader bandwidth would be missed for the same drive and total length. For example, if the detuning step size is 0.1 rad, no stable solutions would be identified at around -650 fs^2, whereas they would with step size 0.01 rad. With finer detuning step sizes (0.01rad), in comparison to Fig. 1, stretched pulses are found at a smaller magnitude of GDD (Fig S3b), with a correspondingly increased bandwidth (Fig S3c). The stretching ratio remains constant but at a higher value (Fig S3d). These results suggest that higher performance may be possible if the locking of the detuning can be achieved with higher precision. This is an interesting direction for future investigation.

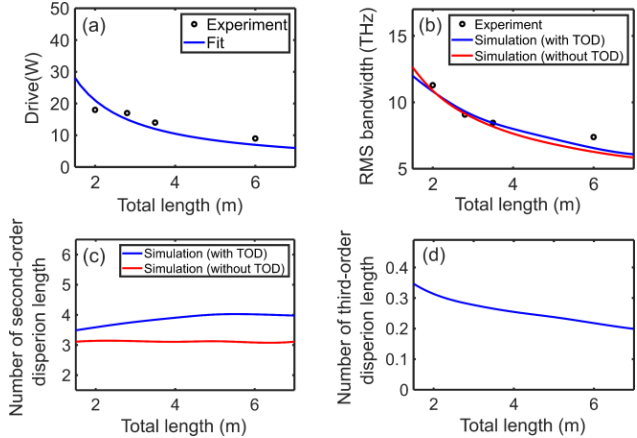

Fig. S2. Stretched-pulse soliton dependence on total cavity length. (a) The max drive power vs. total length from experiments (points) with fit (blue line). (b) The broadest root-mean-square (RMS) bandwidth vs. total length from experiments (points) and simulations with (blue line) and without (red line) TOD. (c) The number of second order dispersion lengths in the cavity at the smallest stable GDD as a function of total cavity length with (blue line) and without (red line) TOD. (d) The number of third order dispersion lengths in the cavity at the smallest stable GDD as a function of total cavity length.

## 4. Autocorrelation measurements

Autocorrelation measurements are used to characterize the temporal profile of the stretched pulse solitons. As discussed in the paper, the fiber after the cavity affects the pulse shape and duration that is output directly from the cavity before measurement. After the cavity output, there is a circulator, fiber-Bragg grating filter, dispersion compensating fiber and an Erbium amplifier (Fig. S4a). Numerically, the intracavity pulse has a duration of 120 fs (Fig S4b blue). After the components succeeding the cavity, the pulse becomes 150 fs, as taken directly before the autocorrelator (Fig S4b red). The autocorrelation of the simulated pulse after the output fibers agrees very well with the experimental measurements (Fig 2 in the main paper), which confirms the generation of 120-fs pulses directly from the fiber Kerr resonator.

## 5. Stretched-pulse trapping analysis

Pulsed driving is used in fiber Kerr resonators to enhance pump peak power [7,8]. In the present experiments, the drive pulses stem from a continuous wave laser modulated into 0.3-ns pulses at a 100-MHz repetition rate amplified by a single-mode erbium-doped fiber amplifier (Fig. S5a). The resulting peak drive power is 18W. The drive pulse is well fit by a 4th order super Gaussian (Fig S5b). The peak drive power can be further improved with a shorter pulse generator.

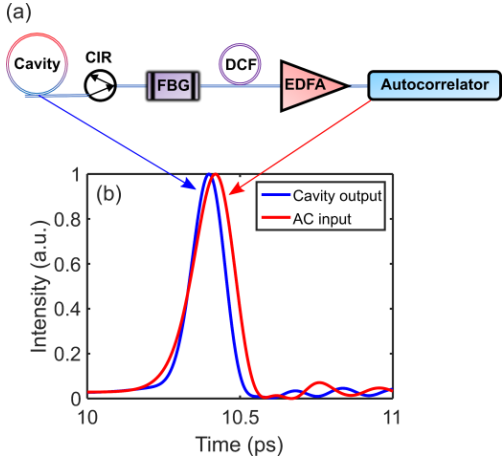

Fig. S4. (a) Schematic of the experimental setup for autocorrelation measurements. CIR, circulator; FBG, fiber Bragg grating; DCF, dispersion compensating fiber; and EDFA, Erbium-doped fiber amplifier. (b) Numerical comparison of stretched-pulse temporal profile directly out of the cavity (blue) and at the autocorrelator (red).

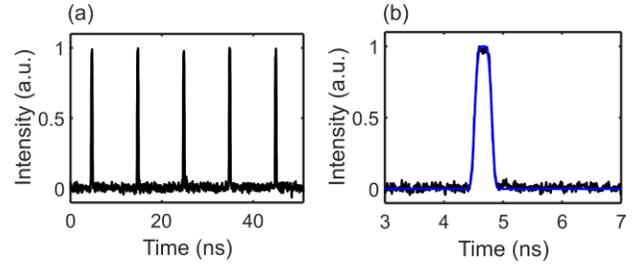

Fig. S5. (a) Drive pulse train measured using a 26-GHz photodetector and a 20-GHz oscilloscope and (b) a single drive pulse (black) with a 4th order Gaussian fit (blue).

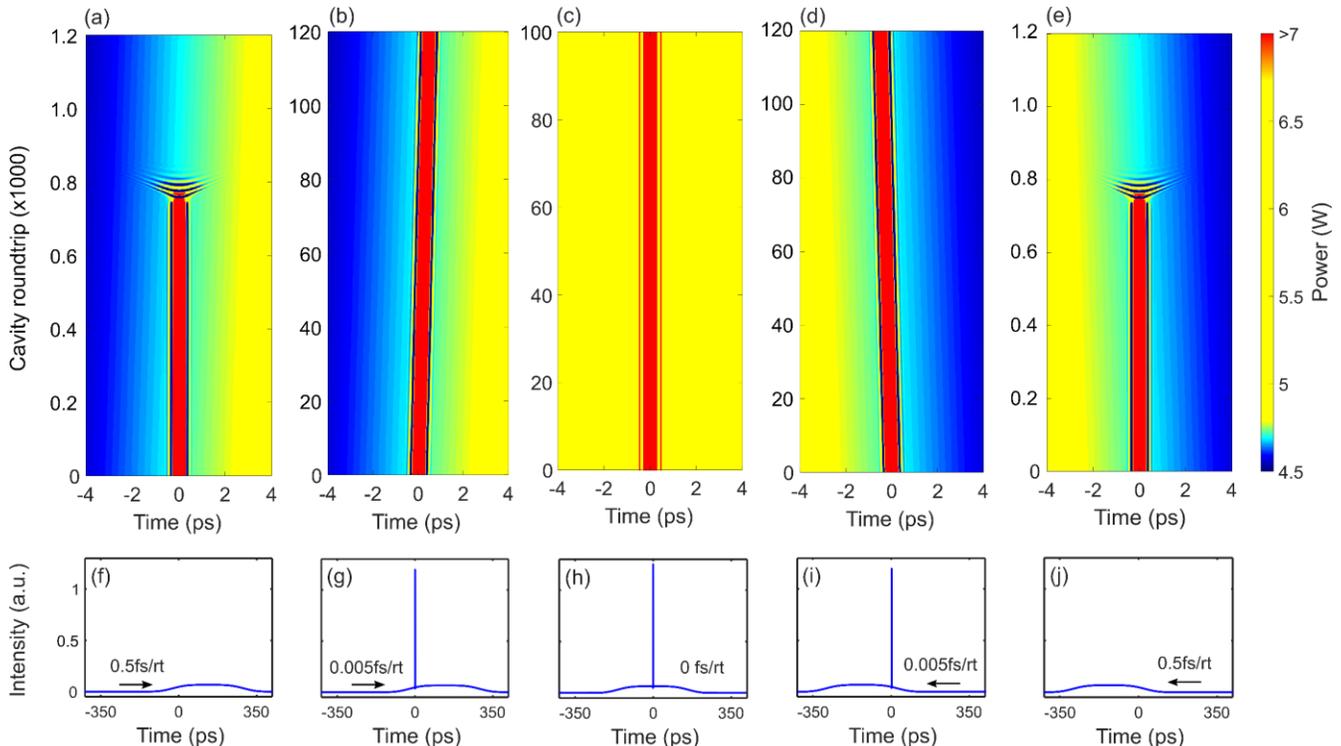

Fig. S6. Numerically simulated pulse evolution for drive-to-soliton desynchronization given by (a), (f) +0.5fs/rt, (b), (g) +0.005fs/rt, (c), (h) 0fs/rt, (d),(i) -0.005fs/rt, and (e), (j) -0.5fs/rt. rt, roundtrip.

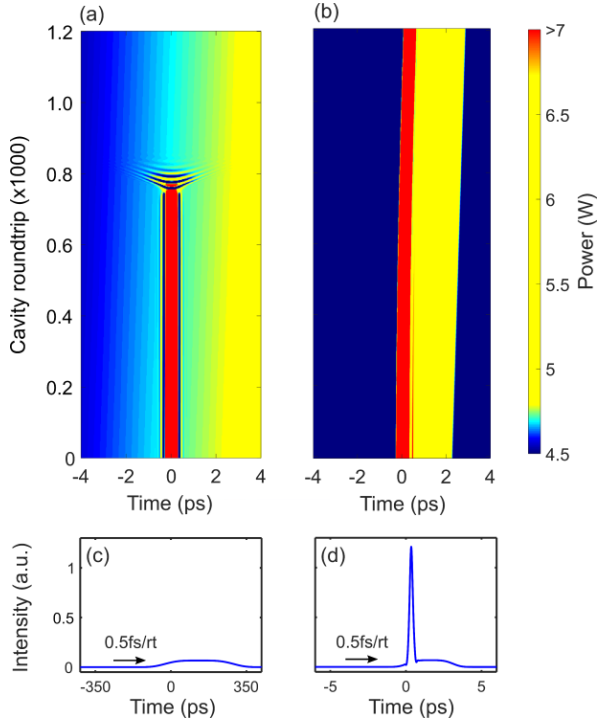

Fig. S7. Numerical simulation with 0.5fs/rt drive-to-soliton desynchronization: (a) Pulse evolution and (c) the final temporal profile for 300-ps drive duration and (b) pulse evolution and (d) the final temporal profile for 3-ps drive duration.

Trapping is investigated as a function of the cavity-drive desynchronization, where desynchronization is defined here as drift in femtoseconds per round trip (fs/rt). When the relative drift is large (e.g. >0.5fs/rt), the drift overpowers the trapping force, the soliton does not trap and is eventually not overlapped with the drive, resulting in its decay (Fig S6a,e). When the drift is sufficiently small (e.g. 0.005fs/rt), the trapping force compensates for the drift, and a stable soliton is trapped (Fig S6b-d). The stretched-pulse soliton is trapped at the edge of the drive pulse where it experiences sufficient trapping force, as in the case for anomalous dispersion solitons [3]. In the trivial case of no drift where trapping is not needed, the soliton is also stable, but at the initial location at the center of the drive pulse (Fig. S6c). When the direction of the drift changes, the trapping location with respect to the drive pulse switches to the other side, but the trapping threshold drift is unchanged. The threshold drift speed is strongly dependent on the slope on the sides of the drive pulse profile, as was the case in prior studies of anomalous dispersion solitons [9]. To examine this property for stretched-pulse solitons numerically, we decrease the drive pulse duration to 3 ps. In this case the pulse was found to trap for a drift of 0.5fs/rt. (Fig S7 b,d), which was not possible with the longer 300-ps drive pulse from the paper, with long rise and fall times (Fig S7a,c). This confirms that the rise and fall times of the pump pulse strongly influences the trapping strength for stretched-pulse solitons in Kerr resonators. The soliton parameters may also affect trapping, which merits further investigation.

## 6. The addressing source and gate

The addressing pulse source for this experiment has 1-nm spectral bandwidth and an 8.2-MHz repetition rate [10] (Fig S8a blue). To achieve the high peak power required for addressing [5], the pulses are amplified before being coupled into the cavity. After the amplifier, in addition to the desired amplified pulse train, there is an additional amplified spontaneous emission background peak at around 1530nm (Fig S8a red). Before exciting solitons in experiments, the addressing beam is attenuated by ~20dB at 1550nm with an intensity modulator (Fig. S8b black). However, despite 20dB of attenuation at 1550nm, the peak at 1530nm remains (Fig. S8b black). When the addressing pulses are not attenuated by the intensity modulator (Fig S8b red), addressing pulses with high peak power are coupled into the cavity and stretched-pulse solitons are excited as detailed in the paper. The residual 1530nm peak can be removed for any practical realization of fiber Kerr resonators for applications.

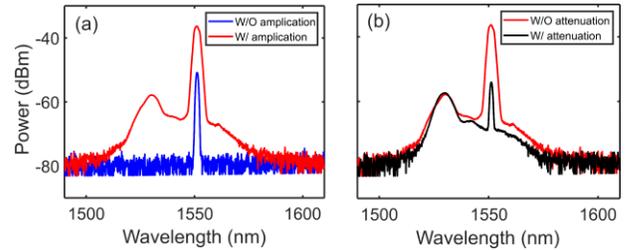

Fig. S8. (a) Spectrum of the addressing pulse with and without amplification (b) Spectrum of the addressing pulses with and without attenuation before coupling into the cavity.

Note that at least one addressing pulse is used in each case to initiate Kerr solitons. After initiation, no further addressing pulses are required if the drive pulse is synchronized with the soliton. Without soliton trapping, a new addressing pulse is needed to periodically reinitiate the soliton when lost due to desynchronization. This is achieved by continuously addressing pulses into the cavity by keeping the intensity modulator gate open. With soliton trapping, the drive is temporally aligned with the soliton, and the soliton only needs to be addressed once by allowing one mode-locked laser pulse to enter the cavity by opening the intensity modulator gate for a period shorter than twice the mode-locked laser period.

## 7. Potential application to microresonators

It will be interesting to investigate how the dependence of performance on total cavity length would scale for microresonators. In microresonators, while the mode area is smaller, the cavity length is also shorter, which amounts to comparable total nonlinearity to fiber cavities. However, the total dispersion of microresonators is much smaller than that of fiber cavities. For stretched pulses with the same spectral bandwidth, the pulses in microresonators will have a much smaller stretching ratio. For example, stretched-pulse solitons have been investigated recently in microresonators with a modest 1.08 stretching ratio [11]. The present study suggests that much shorter pulses can be generated by reducing the cavity net dispersion until it reaches stretching ratios near 3.

### 8. An additional note regarding spectral mode density

In this study we find that by decreasing the cavity length, while the spectral mode density increases linearly with repetition rate, the total bandwidth increases nonlinearly at a slower rate, resulting in an overall decrease in the number of modes supported by the spectrum. However, reducing the number of modes is not required to increase the bandwidth. Decreasing the total dispersion in each fiber is the essential requirement for increasing the bandwidth. Here we achieve this by reducing the total cavity length, which reduces the spectral mode density. However, an alternative is to use fibers with smaller dispersion per unit length, which would increase the bandwidth without reducing the spectral mode density. This approach may be valuable for applications that require high bandwidth and spectral mode density.

## References

1. Y. Wang, M. Anderson, S. Coen, S. G. Murdoch, and M. Erkintalo, "Stimulated Raman Scattering Imposes Fundamental Limits to the Duration and Bandwidth of Temporal Cavity Solitons," Phys. Rev. Lett. **120**, 053902 (2018).
2. C. Milián, A. V. Gorbach, M. Taki, A. V. Yulin, and D. V. Skryabin, "Solitons and frequency combs in silica microring resonators: Interplay of the Raman and higher-order dispersion effects," Phys. Rev. A **92**, 033851 (2015).
3. C. Aguergaray, A. Runge, M. Erkintalo, and N. G. R. Broderick, "Raman-driven destabilization of giant-chirp oscillators: Fundamental limitations to energy scalability," Opt. Lett. **38**, 2644–2646 (2013).
4. L. A. Lugiato and R. Lefever, "Spatial dissipative structures in passive optical systems," Phys. Rev. Lett. **58**, 2209–2211 (1987).
5. F. Leo, S. Coen, P. Kockaert, S. P. Gorza, P. Emplit, and M. Haelterman, "Temporal cavity solitons in one-dimensional Kerr media as bits in an all-optical buffer," Nat. Photonics **4**, 471–476 (2010).
6. M. Haelterman, S. Trillo, and S. Wabnitz, "Dissipative modulation instability in a nonlinear dispersive ring cavity," Opt. Commun. **91**, 401–407 (1992).
7. Y. Xu, A. Sharples, J. Fatome, S. Coen, M. Erkintalo, and S. G. Murdoch, "Frequency comb generation in a pulse-pumped normal dispersion Kerr mini-resonator," Opt. Lett. **46**, 512 (2021).
8. Z. Li, Y. Xu, S. Coen, S. Murdoch, and M. Erkintalo, "Experimental observations of bright dissipative Kerr cavity solitons and their collapsed snaking in a driven resonator with normal dispersion," Optica **7**, 1195 (2020).
9. M. Erkintalo, S. G. Murdoch, and S. Coen, "Phase and Intensity Control of Dissipative Kerr Cavity Solitons," J. R. Soc. New Zeal. 1 (2021).
10. X. Dong, Q. Yang, C. Spiess, V. G. Bucklew, and W. H. Renninger, "Stretched-Pulse Soliton Kerr Resonators," Phys. Rev. Lett. **125**, 33902 (2020).
11. Y. Li, S. Huang, B. Li, H. Liu, J. Yang, A. K. Vinod, K. Wang, M. Yu, D. Kwong, H. Wang, K. K. Wong, and C. W. Wong, "Real-time transition dynamics and stability of chip-scale dispersion-managed frequency microcombs," Light Sci. Appl. **9**, 52 (2020).